\documentclass[aps,prb,twocolumn,preprintnumbers]{revtex4}
\usepackage{graphicx}
\usepackage{amstext}
\usepackage{array}
\usepackage{amssymb}
\usepackage{amsmath}
\usepackage{color}
\newcommand{\be}{\begin{equation}}
\newcommand{\ee}{\end{equation}}
\newcommand{\bea}{\begin{eqnarray}}
\newcommand{\eea}{\end{eqnarray}}
\newcommand{\Eq}[1]{Eq.\,(\ref{#1})}
\newcommand{\Fig}[1]{Fig.\,\ref{#1}}
\newcommand{\Sec}[1]{Sec.\,\ref{#1}}
\newcommand{\Tab}[1]{Table\,\ref{#1}}
\newcommand{\Cite}[1]{~[\onlinecite{#1}]} 

\newcommand{\br}{\mathbf{r}}

\newcommand{\heps}{\hat{\boldsymbol{\varepsilon}}}

\newcommand{\hsigma}{\hat{\boldsymbol{\sigma}}}

\newcommand{\meps}{\varepsilon_j}
\newcommand{\mepsi}{\varepsilon_j''}
\newcommand{\mepsr}{\varepsilon_j'}

\newcommand{\mepst}{\Delta\varepsilon_j}
\newcommand{\mepsrt}{\Delta\varepsilon_j'}
\newcommand{\mepsit}{\Delta\varepsilon_j''}

\newcommand{\eps}{\varepsilon}
\newcommand{\epsi}{\varepsilon''}

\newcommand{\feps}{\varepsilon(\omega_j)}
\newcommand{\fepsi}{\varepsilon''(\omega_j)}
\newcommand{\fepsr}{\varepsilon'(\omega_j)}

\newcommand{\sigmar}[1]{\sigma_{#1}'}
\newcommand{\sigmai}[1]{\sigma_{#1}''}

\begin{document}
\title{Optimizing the Drude-Lorentz model for material permittivity: \\Examples for semiconductors}
\author{H.\,S. Sehmi}
\author{W. Langbein}
\author{E.\,A. Muljarov}
\affiliation{School of Physics and Astronomy, Cardiff University, Cardiff CF24 3AA,
United Kingdom}
\begin{abstract}
Approximating the frequency dispersion of the permittivity of materials with simple analytical functions is of fundamental importance for understanding and modeling their optical properties. Quite generally, the permittivity can be treated in the complex frequency plane as an analytic function having a countable number of simple poles which determine the dispersion of the permittivity, with the pole weights corresponding to generalized conductivities of the medium at these resonances. The resulting Drude-Lorentz model separates the poles at frequencies with zero real part (Ohm's law and Drude poles) from poles with finite real part (Lorentz poles). To find the parameters of such an analytic function, we minimize the error weighted deviation between the model and measured values of the permittivity. We show examples of such optimizations for various semiconductors (Si, GaAs and Ge), for different frequency ranges and up to five pairs of Lorentz poles accounted for in the model.
\end{abstract}
%
%
\date{\today}
\maketitle

\section{Introduction}

In this paper we expand on our recent work\Cite{Sehmi16} on fitting a generalized Drude-Lorentz (DL) model to the experimental permittivity of metals (gold, silver and copper), by applying the same method to common semiconductors. The fitting procedure employs an error minimization using a semi-analytic approach in which we first exactly determine the pole weights for given pole frequencies by solving a set of linear equations. This reduces the number of remaining fit parameters by about a factor of two. These parameters, which are the pole frequencies, are then determined numerically using a gradient decent.

The use of an analytical model of $\heps(\omega)$, which contains only simple poles, is motivated by physical arguments, such as the presence of resonances in the material self energy and response functions. Furthermore, this form of the permittivity can be efficiently implemented in numerical methods, such as the finite difference in time domain (FDTD) method \Cite{VialJPD07}, and in more analytic and rigorous approaches, such as the dispersive resonant-state expansion (RSE)\Cite{MuljarovPRB16}. While real metals are described well by the Drude model and as a result have a dominating contribution of Ohm's law and Drude poles, the permittivity of semiconductors is dictated by the interband transitions which are represented by Lorentz components. We will therefore use here only Lorentz terms for $\heps(\omega)$, and call the model a Lorentz model.

We use an efficient algorithm of fitting experimental data with the Lorentz model with an arbitrary number of Lorentz poles, which takes into account experimental errors if available. This algorithm combines an exact analytical approach for the linear parameters of the model, with a numerical solver for the optimization of the nonlinear parameters of the model. This method, which  is discussed in detail in our previous paper\Cite{Sehmi16}, results in a significant speed-up of the process of finding a complete fit for a set of measured data. We have also developed methods of quickly finding the non-linear parameters with MATLAB's fminunc function through suited selection of start values. We show examples of fitted data for Si, GaAs and Ge, for different frequency ranges and up to 5 pairs of Lorentz poles. We illustrate the resulting pole positions and their weights  in the complex plane to give some physical insight how the model approximates the electronic transitions in real materials.

The paper is organized as follows. \Sec{DL} introduces a generalized DL model of the permittivity and explains why we use in the present work a reduced version of this model. The fit procedure, including the analytical and numerical optimization of parameters of the DL model and the algorithm for determining appropriate starting values in the gradient descent minimization is summarized in \Sec{sec:Opt}. Results of the fit are provided in \Sec{sec:Results} for the semiconductor crystals silicon, gallium arsenide, and germanium.

\section{Lorentz model}
\label{DL}

Quite generally, the permittivity $\heps(\br,\omega)$ can be treated as an analytic function in the complex frequency plane, having a countable number of simple poles. Then, according to Mittag-Leffler theorem, it can be expressed as
\be
\heps(\omega)=\heps_\infty+\sum_j\frac{i\hsigma_j}{\omega-\Omega_j}\,,
\label{eqn:eps}
\ee
where $\heps_\infty$ is the high-frequency value of the permittivity and $\Omega_j$ are the  resonance frequencies -- poles of the permittivity, determining its dispersion, with the weight tensors $\hsigma_j$ corresponding to generalized conductivities of the medium at these resonances. The Lorentz reciprocity theorem requires that all tensors in \Eq{eqn:eps} are symmetric, and the causality principle requires that $\heps(\omega)$ has no poles in the upper half $\omega$ plane and that $\heps^\ast(\omega)=\heps(-\omega^\ast)$\Cite{LandauLifshitzV8Book84}. Therefore, for a physically relevant dispersion, each pole of the permittivity with a positive real part of $\Omega_j$ has a partner at $\Omega_{-j}=-\Omega^\ast_j$ with $\hsigma_{-j}=\hsigma^\ast_j$. Poles with zero real part of $\Omega_j$ have real $\hsigma_j$. For simplicity, we assume in the following an isotropic response, such that the conductivities and thus the permittivity are described by scalars. We note however that it is straightforward to extend the presented treatment to anisotropic response.

We can separate the poles with zero real part of the frequency, which describe the conductivity of materials in the Drude model. There is a pole at zero frequency which represents Ohm's law, corresponding to the $\omega^{-1}$ low-frequency limit of the dispersion. Together with a second pole which has a negative imaginary part, it provides the $\omega^{-2}$ high-frequency asymptotics, originating from the non-zero mass of the charge carriers. In real materials, the carrier mass and the damping can show a frequency dependence, which is not included in the Drude model. To describe such effects, the DC conductivity can be split \Cite{AllenPRB77} into several Drude contributions.

In this work we will describe semiconductors which have a negligible free carrier density, and thus a negligible Drude pole weight. Their susceptibility in the visible and ultraviolet range is dominated by interband transitions. We thus use a version of $\eps(\omega)$ which contains only pairs of Lorentz poles,
\be
\varepsilon(\omega)= \varepsilon_\infty + \sum\limits_{k=1}^L\left(\frac{i\sigma_k}{\omega-\Omega_k} + \frac{i\sigma_k^*}{\omega+\Omega_k^*}\right)\,,
\label{eqn:eps2}
\ee
where $L$ is the number of such pairs. Both the pole positions  $\Omega_k = \Omega'_k+i\Omega''_k$ and  generalized conductivities $\sigma_k = \sigmar{k}+i\sigmai{k}$ are complex. We denote real and imaginary parts of complex quantities with prime and double prime, respectively, and keep using this notation throughout the paper.

The model of the permittivity $\eps(\omega)$ given by the analytic function \Eq{eqn:eps2} with $\Omega''_k\leqslant 0 $ respects the constrain of causality by construction. The parameters of the model, which are the conductivities and the resonance frequencies, have to be determined from the experimentally measured data.

\section{Optimization}
\label{sec:Opt}

With the analytic model \Eq{eqn:eps2} of the permittivity, the task of fitting the experimental data reduces to finding the parameters of the model which minimize the error weighted deviation $E$ between the analytic and the measured values of $\eps$, as this maximizes the probability of the model given the data. Assuming Gaussian errors, we use the squared deviation, weighted with the root-mean-square (RMS) errors:
\be
E = \sum\limits_{j=1}^{N}  \left( \frac{\fepsr - \mepsr}{\mepsrt} \right)^2 + \left( \frac{\fepsi - \mepsi}{\mepsit} \right)^2\,,
\label{eqn:Error}
\ee
where $\eps_j$ are experimental values, $\Delta\eps_j$ are the corresponding errors, and $N$ is the total number of data points taken into account in the fit.
Considering that typical experimental data sets consist of tens to hundreds of points, and $\eps(\omega)$ is an analytic function of $\omega$ with a large number of parameters, typically in the order of a few tens, a robust and efficient algorithm is needed. To achieve this goal, we first make use of an exact, analytical minimization with respect to the parameters in which $\eps$ is linear -- these are all the conductivities and $\eps_\infty$. This is the reason why it is advantageous to fit $\varepsilon$ instead of the complex refractive index $n+i\kappa$, as for the latter none of the parameters is linear. Then for the rest of the parameters, in which $\eps$ is nonlinear -- these are the pole frequencies -- we use an iterative minimization with a gradient decent and a suited selection of starting points.

 To make this linear dependence on parameters more clear, we write the permittivity as
\begin{multline}
\feps=\varepsilon_\infty + \sum_{k=1}^{L}\left[ \sigmar{k}\left(\frac{i}{\omega-\Omega_k} + \frac{i}{\omega+\Omega_k^*}\right) \right.\\
	+ \left. \sigmai{k}\left(\frac{-1}{\omega-\Omega_k} + \frac{1}{\omega+\Omega_k^*}\right) \right]
\label{eps3}
\end{multline}
in which one can see $1+2L$ linear real parameters: $\varepsilon_\infty$, $\sigmar{k}$ and $\sigmai{k}$.
Minimization of the total error $E$, given by \Eq{eqn:Error}, with respect to the linear parameters can be done analytically by setting to zero all the first derivatives of $E$ with respect to these parameters. This provides a set of $1+2L$ linear equations which can be solved numerically exactly using standard linear algebra packages. We can fix the value of $\varepsilon_\infty$ if requested, removing it from the set of linear parameters, by subtracting our chosen value $\varepsilon_\infty$ from $\feps$.

Using the values of the linear parameters determined by exact  minimization of $E$, we now define, via \Eq{eqn:Error}, a new error function $\widetilde{E}$, which has been already minimized with respect to the linear parameters and depends only on the nonlinear parameters, which are the complex frequencies $\Omega_k$ of the Lorentz poles. Overall, there are $2L$ remaining real parameters over which $\widetilde{E}$ has to be minimized.  To represent the average deviation of the model from the measured data points relative to their experimental RMS error, we introduce the relative error of our fit,
\be
S= \sqrt{\frac{\widetilde{E}}{2N}}\,.
\label{S}
\ee

To minimize $\tilde{E}$ over the $2L$ nonlinear parameters we use known minimization algorithms based on the gradient decent (implemented in MATLAB as function `fminunc'). The main challenge is to select suited starting points for the parameters, from which the algorithm finds local minima. The starting points should be selected in a way that the global minimum is amongst the local minima found. Remaining abrupt changes of $S$ with $N$ can, but do not have to, indicate that the global minimum was not yet obtained, and more starting values should be employed.

For $L=1$, we have a pair of Lorentz poles given by a single complex parameter $\Omega_1$. For the starting value of $\Omega_1$, we use a random logarithmic distribution within the range of the measured data, specifically
\be
\Omega_1 = \omega_1 \left( \frac{\omega_N}{\omega_1} \right)^Y - i (\omega_N - \omega_1)N^{Y'-1}
\ee
where $Y$ and $Y'$ are random numbers with a uniform distribution between 0 and 1. The minimization is repeated with different starting points until at least ten resulting $S$ values are equal within 10\%, and the parameters for the lowest $S$ are accepted as global minimum.

\begin{figure}[t]
	\includegraphics[width=\columnwidth]{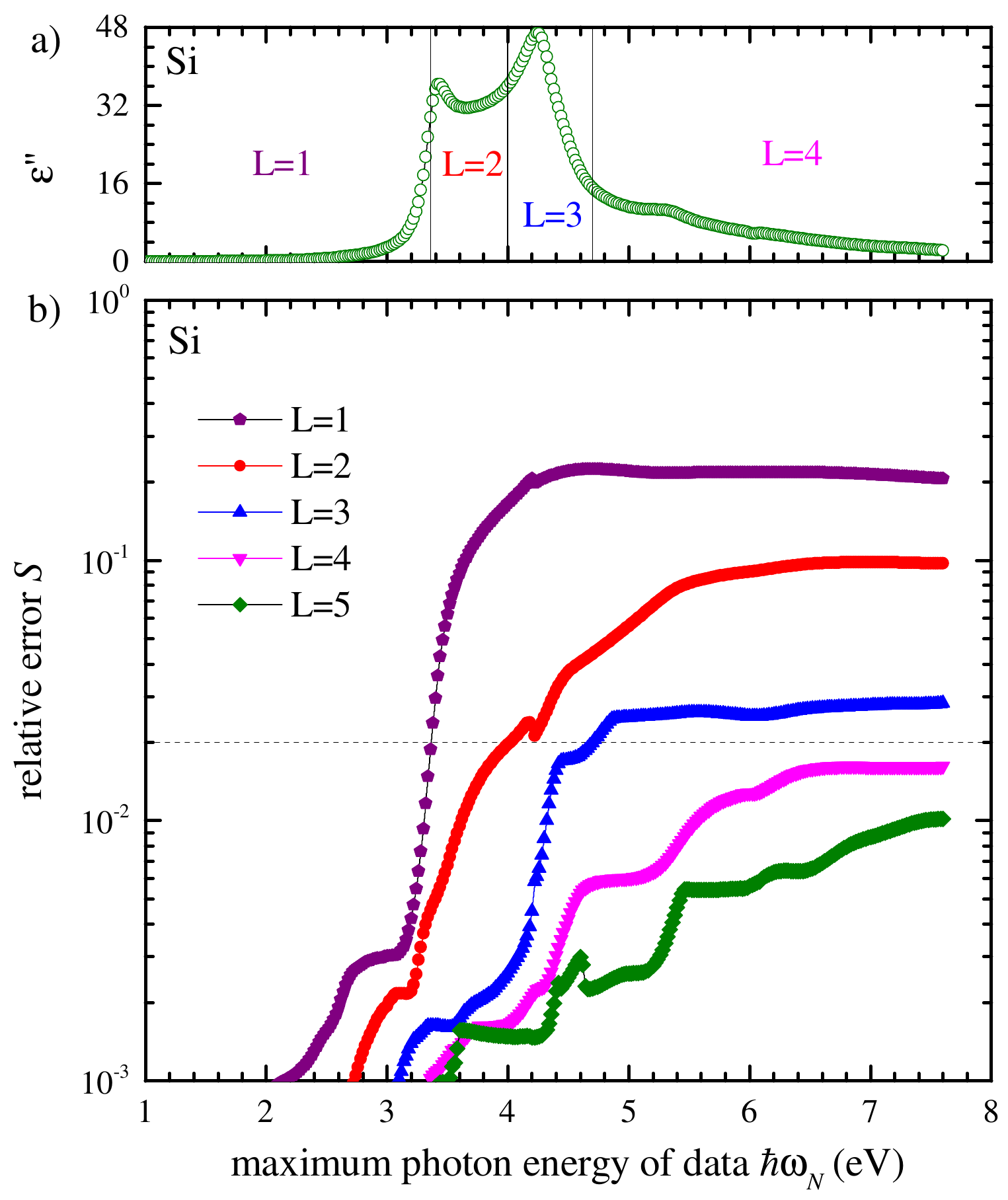}
	\caption{(a) $\mepsi$ as function of $\hbar\omega_j$ for Si. (b) Error $S$ as function of the upper photon energy limit of the fitted data range for Si\Cite{Sopra}. Results for various number of poles in the model are given. Lines are guides for the eye. The maximum photon energy ranges suited for the different  number of poles are indicated in (a) by vertical lines.}
	\label{fig:SiS}
\end{figure}

\begin{figure}[t]
	\includegraphics[width=\columnwidth]{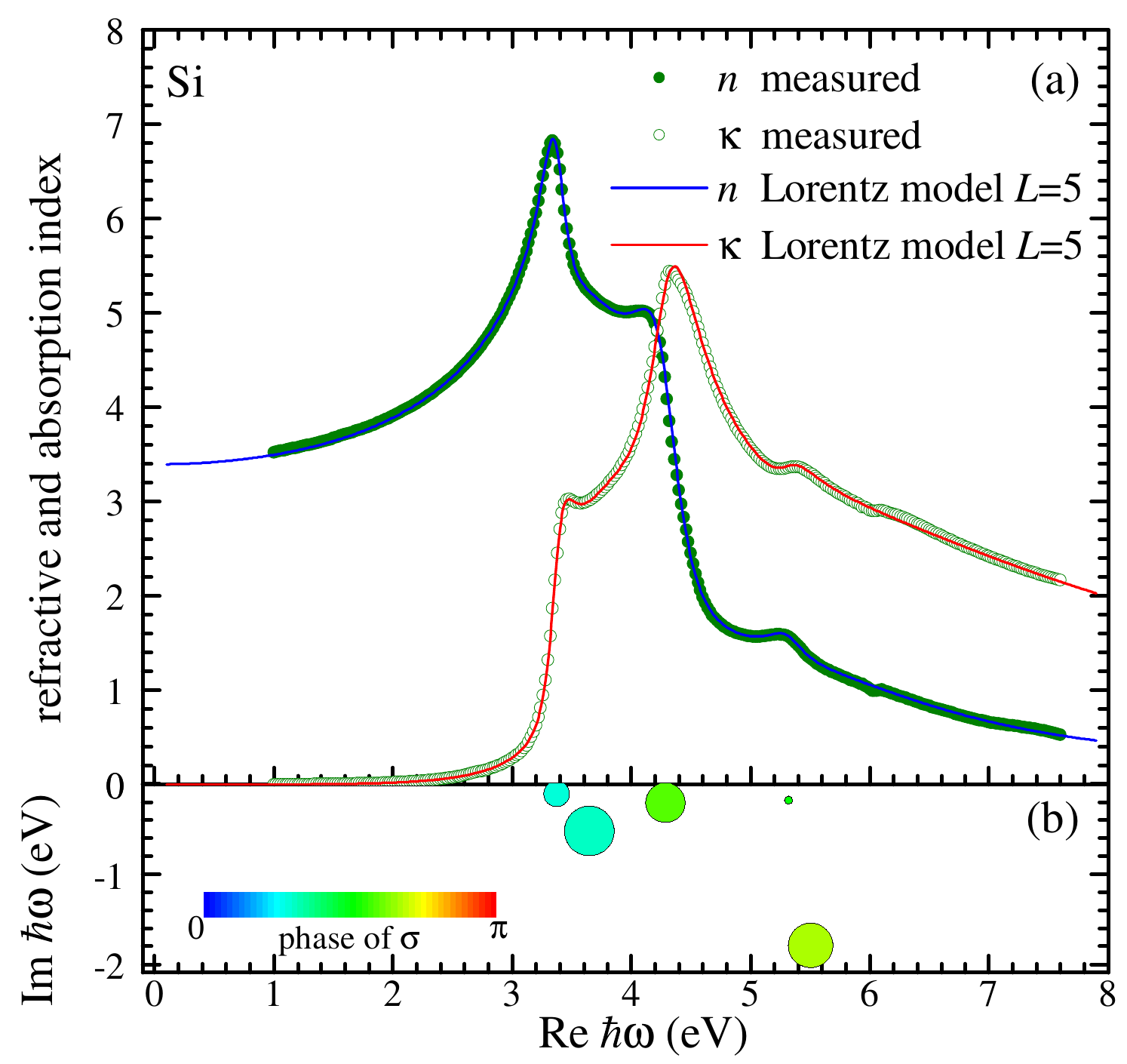}
	\caption{(a) Refractive index $n$ and absorption index $\kappa$ of Si according to\Cite{Sopra} and the Lorentz model \Eq{eqn:eps2} for $L=5$ (solid lines) as functions of the photon energy $\hbar\omega$. The fit is optimized for the full range $1.0\leqslant \hbar\omega \leqslant7.6$\,eV of available data. (b) Pole positions $\Omega_j$ (center of the circle) and weights $\sigma_j$ of the fitted $\eps(\omega)$. The circle area is proportional to $\sqrt{|\sigma_j|}$, and color gives the phase of $\sigma_j$ as indicated. }
	\label{fig:SiL5}
\end{figure}

\begin{figure}[t]
	\includegraphics[width=\columnwidth]{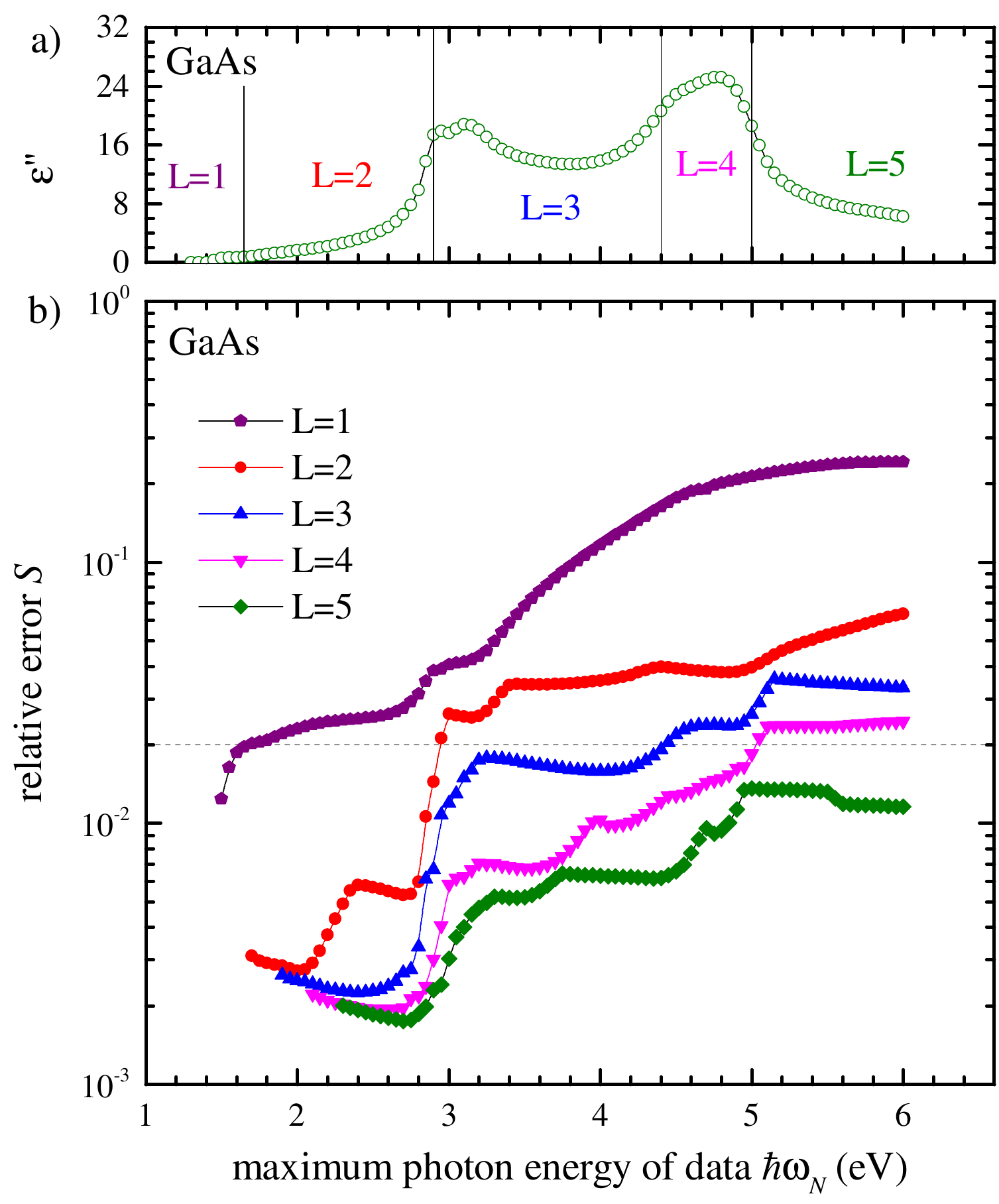}
	\caption{As \Fig{fig:SiS}, but for GaAs.}
	\label{fig:GaAsS}
\end{figure}

\begin{figure}[t]
	\includegraphics[width=\columnwidth]{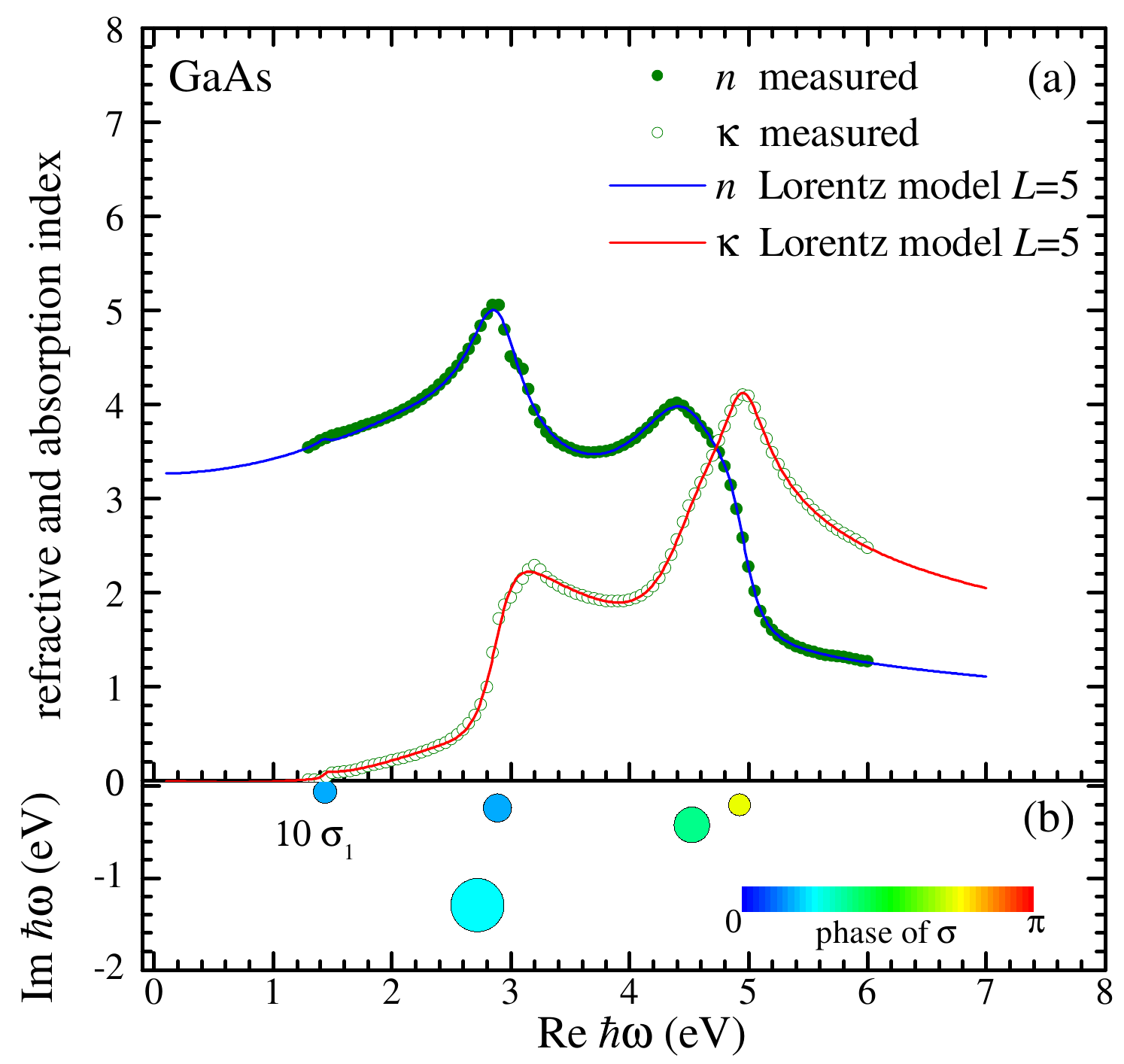}
	\caption{As \Fig{fig:SiL5}, but for GaAs and optimized for the full data range of $\hbar\omega$ given in\Cite{Sopra}, from 1.3\,eV to 6.0\,eV.
}
	\label{fig:GaAsL5}
\end{figure}

The parameter space volume to be covered in such a procedure increases exponentially with $L$, making it computationally prohibitive to use this approach for large $L$. Increasing $L$, we therefore revert to a different strategy. Instead of guessing all $\Omega_k$ randomly, we use the optimized values for  $\Omega_1,...,\Omega_{L-1}$ of the model with $L-1$ poles as starting values for the simulation for $L$ poles, and choose the starting value for the additional pole as $\Omega_L=[1-{i}/{(L+1)}]\omega_N$. This method is fast but can result in not finding the global minimum. However, we can vary the range of the experimental data to be fitted in order to provide more starting points. Here, we choose to keep the lowest frequency $\omega_1$ fixed but vary $\omega_N$ and consequently $N$. Increasing or decreasing $N$ by one, we use as starting point the optimized values for $N$.

Furthermore, going back, from $L+1$ to $L$, just removing one pair of Lorentz poles provides $L+1$ additional starting values for the simulation with $L$ poles. It is also possible to go back multiple steps, e.g., from $L+2$ to $L$ provides $(L+2)(L+1)/2$ starting values -- this however has not been used to produce the $S$ values in this paper.

\section{Results}
\label{sec:Results}

Here we discuss examples of the Lorentz model optimized for measured material dispersions. We show results for crystals of silicon (Si), gallium arsenide (GaAs) and germanium (Ge) using data from\Cite{Sopra}. In our previous work we considered metals (gold, silver and copper) and showed the full range of the experimental permittivity fitted with one Drude pole and three or four pairs of Lorentz poles giving a satisfactory fit. As semiconductors do not have a significant free carrier density, they do not require a Drude pole, so we have fitted the experimental permittivity with the background term ($\varepsilon_\infty$) and up to five pairs of Lorentz poles. We use a variable optimization range, from the lowest measured photon energy $\hbar\omega_1$ to a variable upper boundary of the photon energy $\hbar\omega_N$ taking all available measured values. We show the resulting $S$ values for Si in \Fig{fig:SiS}(b) for different numbers of poles taken into account. We can see that with an increasing number of poles, the error is decreasing, as expected considering the increasing number of parameters. Also, increasing $\hbar\omega_N$ results in larger $S$ values, since a model of a given number of parameters is used to describe an increasing number of data.

\begin{figure}[t]
	\includegraphics[width=\columnwidth]{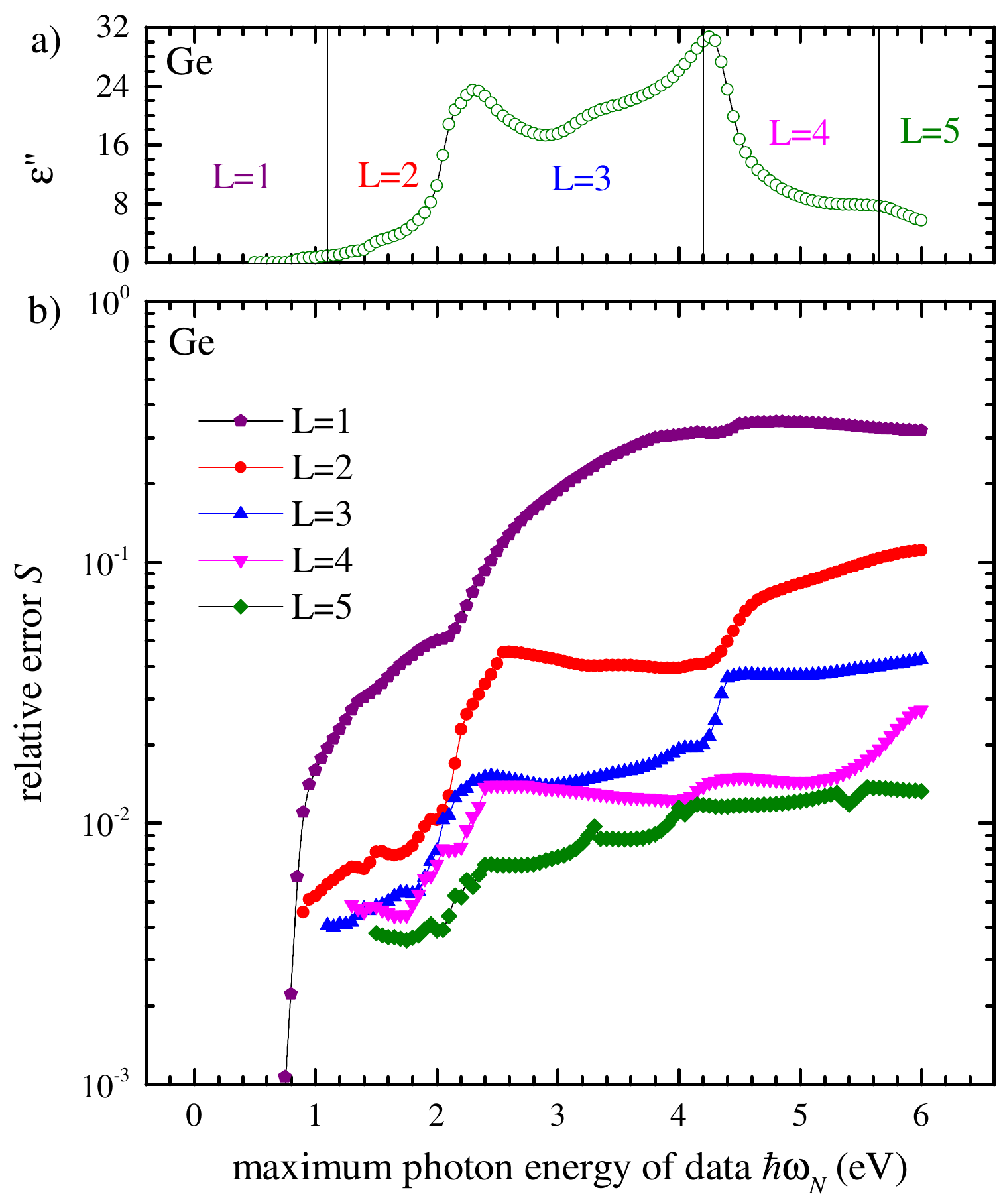}
	\caption{As \Fig{fig:SiS}, but for Ge.}
	\label{fig:GeS}
\end{figure}

\begin{figure}[t ]
	\includegraphics[width=\columnwidth]{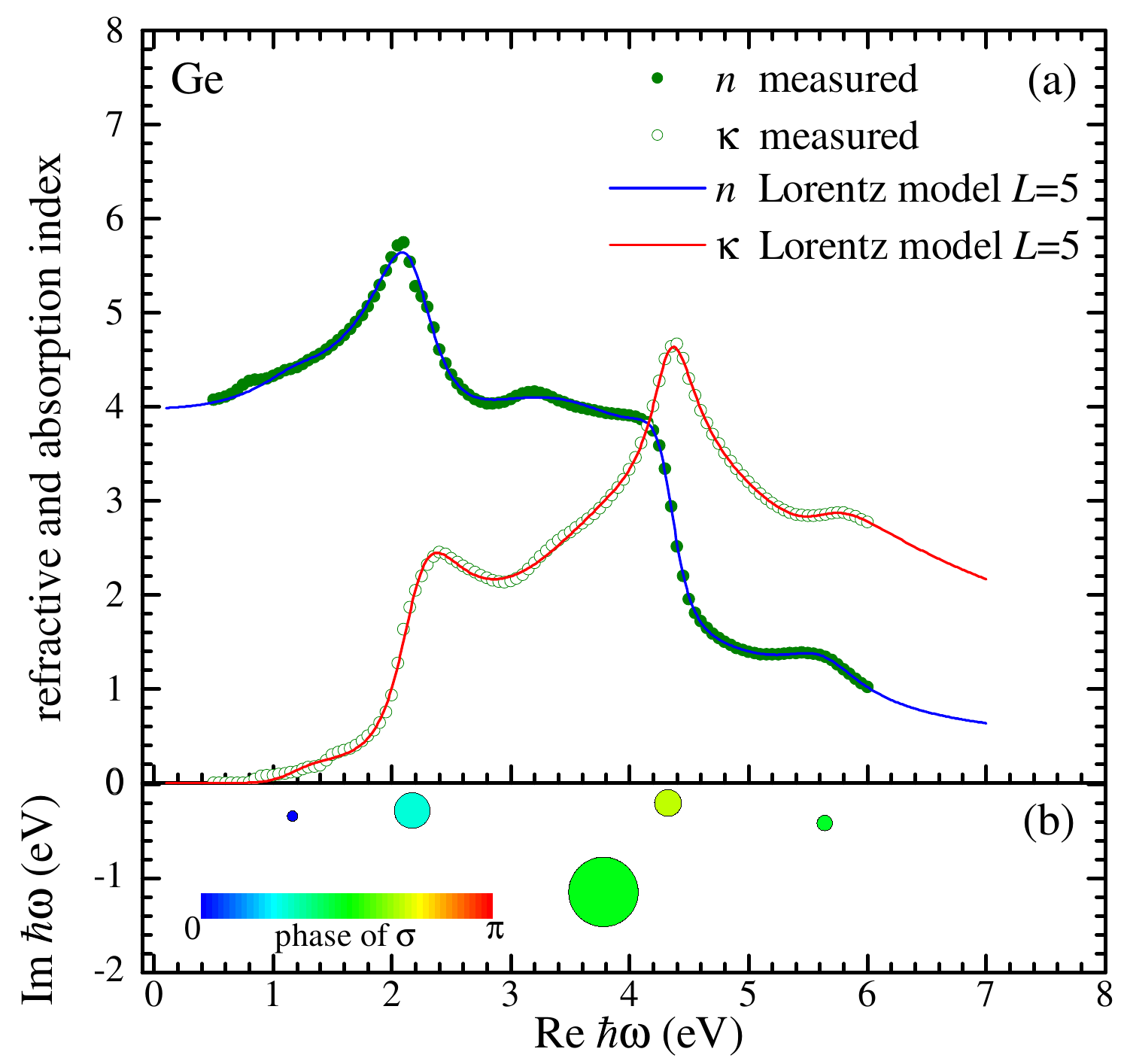}
	\caption{As \Fig{fig:SiL5}, but for Ge and optimized for the full data range of $\hbar\omega$ given in\Cite{Sopra}, from 0.5\,eV to 6.0\,eV.
 }
	\label{fig:GeL5}
\end{figure}

As we have no experimental errors for the permittivity we use in this work $\mepst=\meps$, which means that the $S$ values we use are the normalized relative error. We show a dashed line in \Fig{fig:SiS}(b) at 2\% relative error as a guide to a satisfactory fit.

Concerning the relation of the poles to interband transitions in solids, it is important to emphasize that in microscopic theory the optical response is due to a large number of transitions, often described by a continuum. This continuum, however, can be represented by an infinite or a finite number of poles of the self-energy describing the effects of screening and frequency dispersion. Therefore, the model with a limited number of Lorentz oscillators presents a fully physical though approximate approach, collecting the oscillator strength and transition energies of the continuum into a finite number of poles. The resulting pole positions and weights depend on the energy range to be described and represent microscopic transitions in the material.

For Si, adding the first pair of Lorentz poles ($L=1$) we see the effect of the interband transitions can be described up to the first peak in $\epsi$, around 3.4\,eV. Adding the second pair of Lorentz poles ($L=2$), the effect of the interband transitions can be described up to about 4\,eV, as $\hbar\omega_N$ approaches another peak in $\epsi$. Three pairs ($L=3$) describe  both peaks in $\epsi$ up to 4.7\,eV and finally four pairs of Lorentz poles adequately describes the full range of measured data up to 7.6\,eV, with $S=0.0162$. We do see a significant improvement in $S$ when going to five pairs of poles.

The optimized model for Si with $L=5$ is compared with the experimental data in \Fig{fig:SiL5}(a). The refraction and absorption indices are shown as a function of the photon energy, with the measured data as circles and lines representing the fit functions of the Lorentz model. The poles of the model [see Eq.(\ref{eqn:eps2})] are shown as circles in \Fig{fig:SiL5}(b), centered at their pole positions $\Omega_j$ in the complex photon energy plane, with the complex pole weight represented by the circle area proportional to $\sqrt{|\sigma_j|}$ and the color giving the phase. We find $S=0.0102$ for this fit, with other parameters given in \Tab{tab:para}. We can see that the Lorentz poles are properly positioned to model the interband transitions of silicon. The phases of all 5 poles are close to $\pi/2$, corresponding to a classical damped Lorentz oscillator, such as a mass on a spring. The resonances  $\Omega_1' \sim 3.4$\,eV and $\Omega_3' \sim 4.3$\,eV are around the centers of the two interband transitions well seen in Fig1.(a) within the optimization range, and the half-width of the resonances, $-\Omega_1'' \sim 0.1$\,eV  and $-\Omega_3'' \sim 0.2$\,eV, are approximately covering the half-width of these transitions.

We obtained similar results for GaAs and Ge which can be seen in \Fig{fig:GaAsS} and \Fig{fig:GeS}, respectively. As in \Fig{fig:SiS}, we use a variable upper limit $\hbar\omega_N$ of the optimization range and show the resulting $S$ values for different numbers of poles. For GaAs we find that the model with one pair of Lorentz oscillators works well up to 1.6\,eV, two up to 2.9\,eV, three up to 4.6\,eV, four up to 5\,eV, and five up to a value above the upper limit of 6\,eV.  For Ge we find that the one pair of Lorentz oscillators is a good approximation up to 1.1\,eV, two up to 2.2\,eV, three up to 4.2\,eV, four up to 5.6\,eV, and five beyond the upper limit of 6\,eV.

In \Fig{fig:GaAsL5}, the first Lorentz pole conductivity $\sigma_1$ is multiplied by a factor of 10 for clarity. We find phases of all poles close to $\pi/2$, corresponding to a classical damped Lorentz oscillator. There are similar results for Ge which can be seen in \Fig{fig:GeL5}, most poles having a phase close to $\pi/2$ with the exception of the first pole which has a phase very close to zero. We note that the indirect band gap of both Si and Ge is less suited for the modeling by simple poles, due to the phonon-assisted absorption leading to a weak tail in the absorption spectrum. We can see this clearly in the absorption index for $\hbar\omega<1$\,eV, shown in \Fig{fig:GeL5}(a).

All parameters and values of $S$ for the fits shown in \Fig{fig:SiL5}, \Fig{fig:GaAsL5} and  \Fig{fig:GeL5} are given in \Tab{tab:para}.
\vspace{1cm}

\section{Conclusion}

In conclusion we have demonstrated the performance of our optimization algorithm to determine the parameters of a generalized Drude-Lorentz model for the permittivity of semiconductors, which can be modeled with Lorentz poles only. For $L$ pairs of Lorentz poles taken into account, the developed algorithm uses an analytic minimization over the $2L+1$ linear parameters of the model (the generalized conductivities and high frequency value $\varepsilon_\infty$), and a gradient decent method for determining the $2L$ nonlinear parameters of the model (the Lorentz pole frequencies), with a suited choice of the starting values, resulting in fast and reliable determination of the best global fit. Examples of the fit using up to 5 pairs of Lorentz poles are provided for Si, GaAs and Ge. The optimization program implementing the described algorithm to model any measured data for the refractive index or permittivity is also provided\Cite{Program}.


\acknowledgments This work was supported by the S\^er Cymru National Research Network in Advanced Engineering and Materials.

\begin{table}[t]
  \begin{tabular}{>{$}c<{$} | >{$}c<{$} | >{$}c<{$} | >{$}c<{$}}
	  \text{Material}				& \text{Si}       & \text{GaAs}		& \text{Ge}\\
    \hline
 \rule{0pt}{3ex}
     \varepsilon_\infty	& 0.81568	& -0.54651	& 0.79842\\
   \hline
  \rule{0pt}{3ex}
    \Omega_1' ({\rm eV})	& 3.3736	& 1.4377	& 1.168\\
      \Omega_1'' ({\rm eV})	& -0.11402	& -0.05948	& -0.33778\\
      \sigma_1' ({\rm eV})	& 1.6934	& 0.01981	& 0.47159\\
      \sigma_1'' ({\rm eV})	& 2.084	& 0.01122	& 0.01002\\
   \hline
\rule{0pt}{3ex}
      \Omega_2' ({\rm eV})	& 3.6519	& 2.7229	& 2.174\\
      \Omega_2'' ({\rm eV})	& -0.52378	& -1.2972	& -0.28077\\
      \sigma_2' ({\rm eV})	& 5.2573	& 7.8336	& 3.2926\\
      \sigma_2'' ({\rm eV})   & 8.0106	& 8.3274	& 4.1239\\
   \hline
 \rule{0pt}{3ex}
     \Omega_3' ({\rm eV})	& 4.2877	& 2.8922	& 3.781\\
      \Omega_3'' ({\rm eV})	& -0.21116	& -0.23992	& -1.1461\\
      \sigma_3' ({\rm eV})	& -1.7164	& 2.706	& 0.86584\\
      \sigma_3'' ({\rm eV})	& 5.9939	& 1.616	& 18.898\\
  \hline
\rule{0pt}{3ex}
      \Omega_4' ({\rm eV})	& 5.3188	& 4.5222	& 4.3232\\
      \Omega_4'' ({\rm eV})	& -0.18434	& -0.42072	& -0.20006\\
      \sigma_4' ({\rm eV})	& -0.00528	& 2.1137	& -1.7377\\
      \sigma_4'' ({\rm eV})	& 0.32911	& 4.6445	& 2.5278\\
    \hline
\rule{0pt}{3ex}
      \Omega_5' ({\rm eV})	& 5.5064	& 4.9278	& 5.6442\\
      \Omega_5'' ({\rm eV})	& -1.7892	& -0.19972	& -0.41214\\
      \sigma_5' ({\rm eV})	& -3.8438	& -1.243	& 0.10451\\
      \sigma_5'' ({\rm eV})	& 6.9298	& 1.4424	& 1.0292\\
    \hline
 \rule{0pt}{3ex}
     \hbar\omega_1 ({\rm eV})	& 1.0		& 1.3		&0.5\\
      \hbar\omega_N ({\rm eV})	& 7.6		& 6.0		&6.0\\
      2N					& 662		& 190		&222\\
      1+4L				& 21		& 21		&21\\
      S					& 0.01016	& 0.01157	&0.01327\\
  \end{tabular}
  \caption{Optimized model parameters for different semiconductors, using the fit function with five pairs Lorentz pole and optimization energy ranges corresponding to the data shown in Figs.\,\ref{fig:SiL5}, \ref{fig:GaAsL5}, and \ref{fig:GeL5}. The number of data values $2N$, the number of fit parameters $1+4L$, and the resulting error $S$ are also given.}
  \label{tab:para}
\end{table}

\newpage

\end{document}